\documentclass[12pt]{article}

\usepackage[utf8]{inputenc}
\usepackage[T1]{fontenc}
\usepackage{lmodern}
\usepackage{authblk}

\usepackage[english]{babel}
\usepackage{setspace}
\usepackage[a4paper, margin=2.5cm]{geometry}
\usepackage{parskip}
\usepackage{microtype}

\usepackage[colorlinks=true, linkcolor=blue, citecolor=blue, urlcolor=blue]{hyperref}

\usepackage{amsmath, amssymb, amsfonts}

\usepackage[backend=biber, style=apa,natbib=true]{biblatex}
\usepackage{graphicx}
\usepackage{booktabs}
\usepackage{longtable}
\usepackage{caption}
\usepackage{subcaption}
\usepackage{tabularx}
\usepackage{adjustbox}
\usepackage{tikz}

\usepackage{listings}
\usepackage{xcolor}

\usepackage{lipsum}
\usepackage{pdflscape}
\usepackage{authblk}

\providecommand{\keywords}[1]
{
  \small	
  \textbf{\textit{Keywords---}} #1
}

\newcommand{\nbpapersinitial}{288} 
\newcommand{\nbpapersfilters}{142} 
\newcommand{\nbpapersjournalareas}{125} 
\newcommand{\nbpapersalgo}{24} 
\newcommand{\nbpapersworkshops}{16} 
\newcommand{\conditionnumber}{370}
\newcommand{\kmoscore}{0.815}

\usetikzlibrary{shapes.geometric, arrows.meta, positioning}
\tikzstyle{database} = [cylinder, shape border rotate=90, aspect=0.25, draw, minimum height=1.2cm, text width=2cm, align=center]
\tikzstyle{process} = [rectangle, draw, minimum height=1.2cm, minimum width=4.5cm, text width=4.3cm, align=center]
\tikzstyle{phase} = [draw=none, text width=1cm, align=center]
\tikzstyle{decision} = [diamond, draw, aspect=2, minimum width=3.5cm, minimum height=1.2cm, text width=3.3cm, align=center]
\tikzstyle{arrow} = [thick, -{Stealth}]

\title{An Algorithmic Framework for Systematic Literature Reviews: A Case Study for Financial Narratives}
\date{}
\author[1]{Gabin Taibi}
\author[1]{Joerg Osterrieder}
\affil[1]{University of Twente}

\begin{document}


\maketitle

\begin{abstract}
This paper introduces an algorithmic framework for conducting systematic literature reviews (SLRs), designed to improve efficiency, reproducibility, and selection quality assessment in the literature review process. The proposed method integrates Natural Language Processing (NLP) techniques, clustering algorithms, and interpretability tools to automate and structure the selection and analysis of academic publications. The framework is applied to a case study focused on financial narratives, an emerging area in financial economics that examines how structured accounts of economic events, formed by the convergence of individual interpretations, influence market dynamics and asset prices. Drawing from the Scopus  \footnote{\url{https://www.scopus.com/}} database of peer-reviewed literature, the review highlights research efforts to model financial narratives using various NLP techniques. Results reveal that while advances have been made, the conceptualization of financial narratives remains fragmented, often reduced to sentiment analysis, topic modeling, or their combination, without a unified theoretical framework. The findings underscore the value of more rigorous and dynamic narrative modeling approaches and demonstrate the effectiveness of the proposed algorithmic SLR methodology.
\end{abstract}

\pagebreak


\section{Introduction}
\label{sec:intro}

The influence of narratives on financial markets has recently become a prominent area of study in both economics and finance. The analysis of narrative includes understanding how stories evolve, spread, and impact financial markets over time. By examining the mechanisms through which they form and propagate, researchers aim to uncover their role in shaping expectations, driving investor behavior, and consequently impact market cycles.

In Social Science, narratives are structured accounts or interpretive frameworks through which people understand and act in the social world. The work of \textcite{somers_narrative_1994} advances that they are not simply representations but ontological structures: social life itself is storied. Narratives constitute social identities, guide actions, and embed individuals in relational settings over time and space. These narratives are constituted through emplotment (causal linking of events), selective appropriation, and their temporal and spatial connectivity.

In economics, narratives are studied under the term Narrative Economics, a field focusing on how economic agents use stories to navigate and make sense of complex environments. \textcite{shiller_narrative_2019} argues that economic fluctuations cannot be fully understood through quantitative models alone, as narratives play a critical role in shaping market movements. Thus, \textcite{tuckett_role_2017} highlights that conviction narratives—coherent and emotionally compelling stories—enable investors to act with confidence despite uncertainty, reinforcing collective behaviors in markets and \textcite{shiller_popular_2020} completes this statement describing economic narratives as “stories that offer interpretations of economic events”. We extend this concept by focusing on financial narratives, a subset of economic narratives, which we define as structured interpretations or explanatory frameworks concerning financial markets or economic events, based on available information. Still, the notion of perfect market efficiency has been challenged by \textcite{grossman_impossibility_1980}, who argue that if all information were instantly reflected in prices, investors would have no incentive to acquire costly private information. This critique aligns more closely with the semi-strong form of the Efficient Market Hypothesis, which allows for delayed information diffusion. Furthermore, individual market participants do not perceive and process information uniformly, as their interpretations are shaped by personal expectations and objectives, contextual knowledge, and prior experience, leading them to adopt or reject specific perspectives selectively. Yet, we observe that, over time, these individual interpretations eventually converge into dominant narratives that influence collective beliefs which, in turn, shape financial market behavior.

Despite growing interest in the field, the concept of financial narratives remains underdefined, and the methodologies used to analyze them are diverse and often inconsistent. The purpose of this systematic literature review is to address this gap by offering a clearer picture of how narratives are conceptualized, modeled, and employed in the study of financial markets. In particular, this review aims to clarify how financial narratives are defined and processed across the literature, and assess their reported impact on financial or macroeconomic variables. It also provides a comprehensive overview of the textual analysis techniques employed, with particular attention to the evolution of methods following the introduction of transformer-based architectures. In doing so, the review assesses whether significant methodological advances have been made since early contributions, aiming to identify both conceptual gaps and best practices in this emerging research area. The study addresses the following research questions: \\
(i) How can NLP and textual analysis techniques be used to quantify and model financial narratives? \\
(ii) Can financial narratives modeling enhance financial market dynamics understanding?

More importantly, the relevance of this research lies not only in its theoretical insights but also in its methodological contributions. In fact, the review advances automated knowledge discovery by advanced NLP techniques and machine learning to systematically identify, evaluate, and extract relevant academic literature. This approach contributes to the development of scalable, data-driven techniques for literature synthesis and lays the groundwork for more reproducible research. The proposed framework incorporates transformer-based sentence embedding, dimensionality reduction, and clustering to evaluate relevance against pre-defined research criteria. By integrating these components, the framework not only enables a more systematic and reproducible literature review process, but also improves the assessment of selection quality, ensuring that the selected studies align closely with the research questions. Furthermore, textual analysis is employed to provide researchers with a structured understanding of the remaining publications, thus facilitating the organization and synthesis of findings, as well as the assessment of the thematic relevance and quality of the selected studies.

The review is structured as follows: Section~\ref{sec:methodology} presents the algorithmic framework for systematic literature reviews, including the query design, filtering steps, clustering approach, and quality assessment procedures. Section~\ref{sec:results} applies the methodology to the domain of financial narratives, analyzing research trends, modeling techniques, and conceptual challenges identified in the selected literature. Section~\ref{sec:discussions} discusses the implications of the findings, limitations of existing approaches, and potential directions for future research in financial narrative modeling.


\section{Methodology}
\label{sec:methodology}

Our approach in this literature review is inspired by the work of \textcite{amato_how_2024}, who structured the review process into eight distinct steps: defining the research question, developing and implementing the review methodology, conducting literature exploration and analysis, applying selection criteria (inclusion/exclusion), assessing the quality of selected studies, extracting relevant data, synthesizing findings, and reporting insights. This structured framework, actually based on \textcite{varsha_how_2024}, starts by defining the research problem and formulating clear research questions. A structured review methodology is then developed to guide literature search and analysis. Thus, specific inclusion criteria are applied to filter studies and assess their relevance and quality. Once the final set of papers is selected, key information is extracted and synthesized to generate meaningful insights, organized and presented in a structured report.

\begin{table}[h]
    \centering
    \caption{Summary of article selection criteria.}
    \label{tab:selection_criteria}
    \begin{tabular}{p{10cm} c}
        \toprule
        \textbf{Criteria} & \textbf{Decision} \\
        \midrule
        Inclusion of pre-defined keywords in title, abstract, or keyword list & Inclusion \\
        Article not published in a scientific journal & Exclusion \\
        Article in a language other than English & Exclusion \\
        Article published before 2010 & Exclusion \\
        Duplicates & Exclusion \\
        Low relevance to research questions & Exclusion \\
        Unavailability of the article online & Exclusion \\
        Workshop Proceeds & Exclusion \\
        \bottomrule
    \end{tabular}
\end{table}

However, because the core phase of the current methodology, the study selection, depends heavily on the researcher’s judgment, the overall process remains subjective, difficult to replicate, and prone to bias. To address this, we extend the framework by integrating transformer-based models to automate and refine the selection phase, improving the consistency of inclusion decisions and the assessment of study relevance. This enhancement strengthens quality control and supports a more scalable and reproducible review. As shown in Table~\ref{tab:selection_criteria}, the paper selection process remains fully systematic, following predefined rules. In this section, we present the overall review process summarized in Figure~\ref{fig:paper_selection_diagram}, from the initial research query definition to the data extraction and result analysis phase.

\subsection{Initial research sourcing}

The first phase involves defining the research context and the type of information we aim to extract from the literature. Using Python, this review sources the publications from the Scopus database, specifically retrieving basic information from the results through the Application Programming Interface (API) and scraping the abstract, authors and keywords from Scopus URLs. We also use the same programming language to automatically get the journal ranking and journal information from SCImago Journal Ranking website \footnote{\url{https://www.scimagojr.com/journalrank.php}}.

As this study focuses on narrative modeling using NLP techniques to better understand financial market dynamics, we are particularly interested in literature that addresses both the methodological development of narrative modeling from large textual corpora and the application of NLP—both traditional and recent approaches—in finance. This leads us to formulate two research questions: How can NLP and textual analysis techniques be used to quantify and model financial narratives? Can financial narrative modeling improve our understanding of financial market dynamics? Accordingly, we seek to identify rigorous and relevant academic work across three main themes: theoretical discussions of narratives in finance and economics (financial narratives theory), the use of NLP and text analysis to extract narratives in financial contexts (NLP for narratives processing), and more specifically, the quantification and tracking of narratives for financial applications (financial narratives modeling).

To initiate the selection process, we designed a targeted search query using the advanced search function of the Scopus database. We experimented with various combinations of keywords in an effort to construct a query that is both broad enough to capture relevant research and narrow enough to exclude unrelated domains. The chosen query captures a wide range of pertinent literature, though it also retrieved some irrelevant papers, such as those focused on narratives in health or psychology, where narratives are also a major topic of study. The query was constructed to retrieve publications related to narratives, NLP, and financial markets:
\begin{itemize} 
    \item \textbf{Economic/Financial Narratives in title, abstract or keywords:} \newline TITLE-ABS-KEY("financial narrative" OR "financial narratives" OR "economic narrative" OR "economic narratives")
    \item \textbf{Narratives, NLP/text analysis and financial terms in title:} \newline OR TITLE(narrative* AND ("language processing" OR nlp OR "language understanding" OR nlu OR "text mining" OR "textual analysis" OR "text analysis" OR "text processing" OR lexicon OR sentiment OR "topic modeling" OR "topic extraction" OR "word embedding" OR "word embeddings" OR "entity recognition" OR "narrative processing" OR "narrative modeling") AND (economic* OR macroeconomic* OR "financial market" OR "financial markets" OR "market dynamics" OR "market movements" OR "price dynamics" OR "price movements" OR "financial forecasting" OR "stock markets" OR "equity markets" OR "foreign exchange" OR commodit* OR bond* OR "asset prices" OR "price returns" OR "asset returns" OR "market returns" OR volatility OR "risk management" OR "portfolio management"))
    \item \textbf{Narratives, NLP/text analysis and financial terms in keywords:} \newline OR KEY(narrative* AND ("language processing" OR nlp OR "language understanding" OR nlu OR "text mining" OR "textual analysis" OR "text analysis" OR "text processing" OR lexicon OR sentiment OR "topic modeling" OR "topic extraction" OR "word embedding" OR "word embeddings" OR "entity recognition" OR "narrative processing" OR "narrative modeling") AND (economic* OR macroeconomic* OR "financial market" OR "financial markets" OR "market dynamics" OR "market movements" OR "price dynamics" OR "price movements" OR "financial forecasting" OR "stock markets" OR "equity markets" OR "foreign exchange" OR commodit* OR bond* OR "asset prices" OR "price returns" OR "asset returns" OR "market returns" OR volatility OR "risk management" OR "portfolio management"))Explainability" OR "Counterfactual Explainability" ) )
\end{itemize}
The inclusion of the term “narrative(s)” is central, as it ensures thematic relevance to our research focus. We prioritized papers that explicitly mention “financial narrative(s)” or “economic narrative(s)” in the title, abstract, or keywords. Additionally, to broaden the scope while maintaining relevance, we formulated a more inclusive search strategy targeting titles and keywords (but not abstracts because they tend to be more descriptive and often mention broader methodological or contextual terms that may not reflect the core focus of the paper), combining “narrative(s)” with terms related to text analysis or NLP, and with terms referring to financial markets.

The second phase involves applying additional filters to refine the search and ensure greater homogeneity in the retrieved literature, and remove duplicates entries. In our case, we extended the initial query by restricting the publication year to 2010 or later, limiting the results to peer-reviewed journal articles, and including only English-language publications, resulting in \nbpapersfilters{} papers:
\begin{itemize} 
    \item \textbf{Economic/Financial Narratives in title, abstract or keywords:} \newline Identical to phase 1
    \item \textbf{Narratives, NLP/text analysis and financial terms in title:} \newline Identical to phase 1
    \item \textbf{Narratives, NLP/text analysis and financial terms in keywords:} \newline Identical to phase 1
    \item \textbf{Time constraint:} \newline AND PUBYEAR > 2010
    \item \textbf{Language constraint:} \newline AND LANGUAGE(english)
    \item \textbf{Article constraint:} \newline AND DOCTYPE(ar)
\end{itemize}
At this stage of the selection process, an additional filtering criterion was introduced to enhance thematic alignment. Specifically, journal subject areas were cross-referenced with disciplinary classifications from the SCI journal ranking database to identify and remove publications originating from fields clearly unrelated to the research context. Journals primarily associated with disciplines outside the intended scope were excluded to eliminate noise and improve the quality of the dataset for subsequent phases. Conversely, journals whose subject areas could plausibly intersect with the research objectives, or for which thematic relevance could not be reliably assessed at this stage, were retained for further evaluation. This approach allowed for the early elimination of clearly irrelevant publications while preserving potentially relevant studies for more detailed analysis in later stages of the selection pipeline.

\subsection{Algorithmic selection framework}

This section presents the third phase, consisting in an algorithmic process used to filter research papers based on their relevance. The selection framework leverages NLP and machine learning techniques to classify papers into three categories: high, medium, and low relevance. The process consists of three main stages: textual analysis, dimensionality reduction, and clustering.

\subsubsection{Textual analysis: research properties statements}

The selection methodology relies on a statement similarity approach, in which the user defines key statements that describe the criteria that selected papers should meet. In our case, the statements respectively reflect the research focus, research type, context, methodology, data sources, and research questions:
\begin{itemize}
    \item ``The research discusses Financial Narrative Processing, Financial Narrative Modeling or the use of textual data to understand financial markets'';
    \item ``The research is highly relevant in the context of financial markets, including: equities, foreign exchange, cryptocurrencies, bonds, commodities, or real estate'';
    \item ``The research is a empirical study showcasing the use of textual data to model narratives and understand financial markets dynamics'';
    \item ``The research methodologies include: textual analysis, text mining or Natural Language Processing techniques such as topic modeling, emotion analysis, sentiment analysis, word embeddings or transformer-based models'';
    \item ``The research leverages the following data: large textual datasets, including financial reports, news articles, social media posts, audio or video transcripts, or any other form of financial textual data'';
    \item ``The research helps answering the research question(s): How can NLP and textual analysis techniques be used to quantify and model financial narratives? Can financial narratives modeling enhance financial market dynamics understanding?''
\end{itemize}

These statements define the inclusion criteria for selecting relevant studies, ensuring that only research focused on financial narrative modeling, NLP-based textual analysis, and their impact on financial markets is considered.  

To evaluate how well each paper aligns with the research statements, two transformer-based sentence embedding models were tested. The first is \texttt{intfloat/multilingual-e5-large-instruct}, an open-source model designed for multilingual tasks \parencite{wang_multilingual_2024} and implemented using the HuggingFace SentenceTransformer\footnote{\url{https://huggingface.co/sentence-transformers}} library in Python. The second is OpenAI’s proprietary \texttt{text-embedding-3-small}, also implemented with the Python API library. While \texttt{multilingual-e5-large-instruct} offers the advantage of being open-source and lightweight in terms of computational requirements, OpenAI’s \texttt{text-embedding-3-small} demonstrates better performance in terms of semantic relevance and alignment accuracy, and is therefore selected for the study.

Rather than embedding the six statements directly, we use the chat completion capabilities of OpenAI’s API to generate five close paraphrases for each statement, reducing sensitivity to specific wording. Each set of six paraphrases (the original plus five variations) is embedded individually, and their mean vector is computed to obtain a single representative embedding per statement. In parallel, we concatenate the title, abstract, and keywords of each paper into a single textual input and generate embeddings for these as well. The cosine similarity between each paper embedding and each of the six statement embeddings is then computed, resulting in six similarity scores per paper. Finally, the average of these six scores is computed to represent the paper’s overall relevance score.

\subsubsection{Data preparation}

Before proceeding with the clustering step, the matrix of similarity scores is Z-score standardized for each of the six dimensions to ensure each dimension contributes equally to the clustering algorithm. The next step is to reduce the dimensionality of the data while preserving its most informative components, in the case where we observe high correlation between relevance features. This is achieved using Principal Component Analysis (PCA), which transforms the six-dimensional similarity space into a lower-dimensional representation by identifying the most important variance-explaining components.

We use the Kaiser-Meyer-Olkin (KMO) score—which assesses the adequacy of the data for factor analysis by measuring sampling adequacy—and the Condition Number (CN), which evaluates multicollinearity by measuring the sensitivity of a matrix to numerical inversion. These two indicators guide our decision on whether to apply PCA. Specifically, if the KMO score is below 0.5, PCA is not recommended due to poor sampling adequacy. If the KMO score exceeds 0.7, PCA is considered appropriate. If the KMO score falls between 0.5 and 0.7, we additionally check the CN: if it is greater than 100, indicating severe multicollinearity, PCA is applied; otherwise, it is not. We therefore apply PCA and retain the number of components that together explain 99\% of the variance. This lower-dimensional representation enhances the efficiency of the subsequent clustering process by eliminating noise and redundancy. Papers that initially shared similar similarity profiles across multiple research statements are now grouped based on their principal components, which encapsulate the most distinguishing features of their textual content. This transformation should refine the classification process, ensuring that only the most relevant dimensions contribute to the selection framework. 

\subsubsection{Research clustering}

After reducing the dimensionality of the similarity scores, the next step in the selection process involves clustering the papers into distinct relevance groups. This classification is performed using three clustering methods: K-means, Gaussian Mixture Models (GMM), and Agglomerative Clustering (AC). These methods were selected for their flexibility and their ability to explicitly control the number of output clusters, which is essential for our relevance-based filtering. K-means partitions the data into a fixed number of clusters by minimizing the within-cluster variance, assigning each point to the nearest cluster centroid. Gaussian Mixture Models extend this approach by assuming that the data is generated from a mixture of Gaussian distributions, allowing for probabilistic cluster assignments and more flexibility in capturing ellipsoidal shapes in the data. Agglomerative Clustering is a hierarchical method that builds clusters by iteratively merging the closest pairs based on a linkage criterion, producing a dendrogram from which a desired number of clusters can be extracted. The clustering is performed in the four-dimensional space obtained from the PCA transformation.

Given the diversity in methodologies, objectives, and data sources within the dataset, we adopted a three-cluster approach to categorize papers into high, medium, and low relevance. This method provides a more granular classification, where high-relevance papers strongly align with the systematic review criteria, low-relevance papers are largely unrelated, and medium-relevance papers share some relevant characteristics and require further evaluation. The medium-relevance group can be manually reviewed to refine the selection, but we decided to opt for a single high-relevance cluster and remove the medium- and low-relevance ones because manual review would reintroduce subjectivity into the selection process, undermining the systematic and reproducible nature of the proposed framework. Moreover, our initial search queried a large number of papers with diverse methodologies, objectives, and data sources. Given this heterogeneity, retaining only the high-relevance category ensures that the final selection consists of studies that closely align with the research objectives while minimizing the inclusion of marginally relevant literature, improving both the focus and quality of the systematic review.

We compare the results of each clustering method based on three main evaluation criteria computed on the high-relevance cluster: the average relevance score, the Silhouette score, and the number of papers retained. The Silhouette score measures how well each sample fits within its assigned cluster, balancing cohesion (similarity within the cluster) and separation (dissimilarity with other clusters); higher values indicate more distinct and well-separated clusters. The objective is to maximize all three metrics to ensure that the high-relevance cluster is both coherent and substantial in size. To identify the most suitable method, we computed a composite score by standardizing each metric (Z-score scaling) and assigning weights of 50\% to the average relevance score, 20\% to the Silhouette score, and 30\% to the number of papers.

\subsection{Data cleaning and extraction}

Once the algorithmic selection process was finalized, a final manual validation phase was carried out to refine the dataset and ensure its quality and consistency. This stage aimed to address limitations that algorithmic methods alone could not fully resolve. Specifically, publications whose full text was not accessible, either because of paywall restrictions or the lack of institutional access, were excluded from the corpus. The inability to review these documents in their entirety prevented the assessment of their thematic relevance and methodological quality, thus warranting their removal. In addition, unconventional documents such as workshop proceedings and compilation volumes were identified and excluded. These types of publications often aggregate a large number of individual studies under a general theme, without guaranteeing a consistent focus aligned with the research objectives. Moreover, the scale and heterogeneity of such documents made them impractical to process within the structured framework of this study. This final validation step ensured that the remaining corpus was both thematically coherent and methodologically appropriate for the subsequent literature analysis.

To structure the extraction process, a reporting framework was designed to capture key aspects of each study. This framework includes details on the research purpose, methodology, data sources, dataset characteristics, main findings, and practical implications, the result of which is summarized in appendix (Table~\ref{tab:data_extraction_table}). The purpose of this step is to document how each study contributes to the understanding of financial narratives, particularly in terms of narrative modeling techniques, data sources, and the integration of NLP methodologies. While the current extraction process remains manual, future work will explore the possibility of automating parts of the extraction using additional NLP frameworks. Automation could improve efficiency and scalability, allowing for the rapid processing of a larger number of studies.


\section{Results}
\label{sec:results}

This section presents the key results of our systematic review, dividing the selected contributions into two complementary strands: papers focusing on narrative understanding and those focused on narrative modeling. The distinction lies in the epistemic and methodological aim. Narrative understanding papers aim to conceptualize or interpret the nature and role of narratives in economic contexts, often rooted in theory or qualitative methods. Narrative modeling papers, on the other hand, propose empirical or algorithmic techniques to quantify, extract, or use narratives for forecasting or explanatory purposes.

\subsection{Selection Phase Results}

The paper selection process followed the four-phase pipeline presented earlier, ensuring thematic relevance and methodological rigor. In the first phase, the initial query was applied to retrieve \nbpapersinitial{} publications without preliminary restrictions on date or subject area, maximizing the initial recall of potentially relevant studies. In the second phase, a series of metadata-based filters were applied. Publications were restricted to journal articles written in English and dated after 2010. To eliminate papers clearly outside the financial or economic domains, we automatically excluded publications that appeared in journals associated with fields unrelated to the topic. These were manually identified by subject areas such as 'Arts and Humanities', 'Medicine', 'Health Professions', 'Earth and Planetary Sciences', 'Environmental Science', 'Agricultural and Biological Sciences', 'Biochemistry, Genetics and Molecular Biology', and 'Energy'. However, we chose not to exclude papers from journals listed as 'Engineering', 'Social Sciences', 'Multidisciplinary', or with no assigned category, as their relevance is evaluated during later stages of the selection process. Duplicate entries were also identified and removed during this phase, reducing the corpus to \nbpapersjournalareas{} publications.

The third phase introduced the algorithmic selection framework. Transformer-based sentence embeddings were computed for the title, abstract, and keywords of each paper. In our use-case, the KMO score is \kmoscore{}, indicating strong sampling adequacy, and the CN is \conditionnumber{}, which further confirms significant multicollinearity. Both indicators suggest that PCA is well suited for this dataset and we therefore obtained four components for use in subsequent analysis. Various clustering methods were evaluated: for K-means, GMM, and AC, we obtained average relevance scores of 0.507, 0.503, and 0.481; Silhouette scores of 0.352, 0.288, and 0.403; and retained paper counts of 24, 22, and 50, respectively. Based on the weighted composite score, K-means was selected as the most appropriate clustering method, resulting in a final set of 24 papers in the high-relevance group.

In the final phase, the manual validation process, two papers were found to be inaccessible, and several workshop proceedings were identified among the selected documents. Specifically, four workshop papers were detected, each containing more than one hundred individual studies. Given the uncertainty regarding their thematic alignment with the research objective and the impracticality of processing them individually within the scope of this review, these documents were excluded from the final dataset. After this refinement step, a total of \nbpapersworkshops{} papers remained for detailed analysis. The evolution of the paper count through each phase of the process is summarized in appendix (Figure~\ref{fig:paper_selection_diagram}).

\begin{figure}[h]
    \centering
    \includegraphics[width=0.8\textwidth]{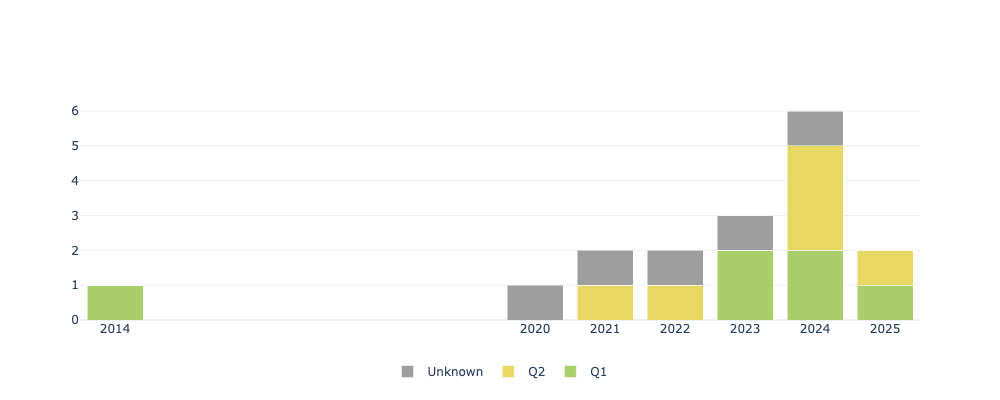}
    \caption{Temporal distributions of selected research papers.}
    \label{fig:paper_temporal_distribution}
\end{figure}

As illustrated on Figure~\ref{fig:paper_temporal_distribution}, the temporal distribution of the final selected research spans from 2014 to 2025, with a majority of studies published in recent years. This distribution suggests a growing interest in the application of NLP techniques to financial narrative analysis, particularly in the last five years. The papers originate from journals covering multiple academic disciplines, including 'Economics, Econometrics and Finance', 'Social Sciences', 'Business, Management and Accounting', 'Computer Science', 'Mathematics', 'Psychology', and 'Decision Sciences'. The ranking of the journals, obtained from the SJR website, reveals that 6 papers were published in Q1 journals, also 6 in Q2 journals, and 5 papers published in journals that do not have a ranking.

\subsection{Narrative Understanding Papers}

The exploration of narrative understanding in financial contexts has grown significantly, particularly as researchers began to recognize that financial narratives go beyond surface-level sentiment or topic extraction. Early works laid a foundation for understanding how narratives shape belief systems, institutional expectations, and investor behavior.

\textcite{hu_annotation_2021} developed a detailed annotation model to detect opinion and emotion expressions in economic texts. Their methodology, which integrates intra-sentential labeling with economic terminology and rhetorical patterns, emphasizes the importance of understanding how subjective language conveys implicit judgments and anticipatory assessments in financial contexts. The resulting corpus, drawn from central bank speeches, 10-k fillings, news, and social media, allows for fine-grained linguistic modeling of investor and institutional sentiment, offering a window into how narratives materialize from text. From a multilingual and applied perspective, \textcite{zmandar_cofif_2022} introduced CoFiF Plus, the first large-scale corpus of French financial narrative summaries. Though the paper’s primary aim was dataset construction, the authors contextualized financial narratives as communicative devices used by firms to shape investor perceptions. Their discussion reinforces the idea that narratives are vehicles for trust and persuasion, embedded in institutional and linguistic norms. \textcite{sy_fine-grained_2023} also proposed a fine-grained argument mining approach applied to financial earnings calls, using BERT ensembles to classify and relate argumentative units. Their focus on logic structures and discourse coherence highlights the interpretative structure of financial narratives, especially in how sentiment and logic intertwine in decision-relevant communication.

\textcite{liu_beyond_2024} addressed the subtlety of year-over-year semantic drift in financial reports. They introduced the Financial-STS task to quantify nuanced differences in narratives from company disclosures. Their work moves beyond superficial textual similarity, proposing a categorization of semantic shifts (e.g., intensified sentiment, emerging situations) that reflect how managerial narratives evolve over time. This conceptual framework offers a way to systematically detect narrative shifts relevant for investors.

Finally, \textcite{roos_narratives_2024} took a more theoretical stance, arguing that much of the empirical literature confuses topics with narratives. They provided a rigorous definition of collective economic narratives as socially shared, action-oriented stories that emerge in context and suggest coordinated beliefs or behaviors. Their contribution lies in reconciling insights from institutional economics, literary theory, and complexity economics to establish conceptual clarity. Importantly, they emphasize that narratives are distinct from mere themes or sentiment—they must be sense-making and socially transmitted.

Together, these contributions emphasize that financial narratives are not simply collections of words or topics, but structured, evaluative, and often persuasive expressions grounded in economic language, argumentation, and context. The understanding of narratives therefore requires both conceptual clarity and methodological precision—ranging from theoretical definitions to corpus-level annotation and semantic modeling.

\subsection{Narrative Modeling Papers}

The second group of studies focuses on modeling narratives through quantitative methods, aiming to extract, represent, and utilize narrative signals in financial prediction or macroeconomic analysis. These works share a common methodological objective: translating textual narratives into structured, predictive signals for the economy or financial markets.

Among the earliest contributions, \textcite{tuckett_tracking_2014} proposed a social-psychological framework grounded in conviction narrative theory, analyzing the dynamics of emotionally charged narratives---termed ``phantastic objects''---and their impact on financial decision-making. Using sentiment shift detection on Reuters articles and the Enron email archive, they developed a method to capture rising emotional conviction and eventual disillusionment in narrative content, offering one of the first systematic approaches to narrative regime detection in financial text.

\textcite{hsu_narrative_2021} explored how narrative salience in historical Chinese news sources related to the U.S. Silver Purchase Act could forecast economic indicators such as price levels. This study applied textual frequency analysis combined with regularized regression, highlighting the value of narrative-based proxies even in early 20th-century macroeconomic settings.

In contemporary financial markets, \textcite{chen_covid_2022} extracted narrative features related to the COVID-19 pandemic and studied their influence on financial volatility. They showed that high narrative virality and pessimism coincided with extreme market movements, demonstrating the real-time utility of narrative indicators in turbulent contexts.

\textcite{zhu_sentiment_2023} applied deep learning and natural language processing on social media data to model housing market sentiment in China. They built forward- and backward-looking indices using LSTM-based models trained on real estate-related posts, showing that narrative sentiment predicted price dynamics and investor expectations in housing.

\textcite{borup_quantifying_2023} collected free-text investor expectations from U.S. households and applied Latent Dirichlet Allocation (LDA) topic modeling to extract narratives. Their results revealed that these subjective narrative features significantly predicted excess asset returns, outperforming traditional sentiment indices. \textcite{ma_stock_2024} also used LDA on a Narrative-based Energy General Index (NEG) constructed with news from Wall Street Journal articles, specifically targeting energy-related topics. Their results showed that NEG predicted returns in the energy sector and outperformed both macroeconomic and sentiment-based predictors. Moreover, NEG showed predictive power across multiple sectors and the aggregate market, offering a robust narrative signal for asset allocation.

From an emotional angle, \textcite{agarwal_investor_2024} used dictionary-based emotion metrics to study Chinese stock bubbles. They found that emotions such as excitement and anxiety embedded in media narratives had strong explanatory power for stock returns, trading volume, and volatility, suggesting that financial cycles are heavily narrative-driven.

\textcite{miori_narratives_2023} applied large language models (GPT-3.5) and graph theory to construct weekly networks of news narratives. Their findings indicated that fragmentation in these narrative graphs often preceded market dislocations, proposing a novel structural indicator of systemic risk. \textcite{stander_news_2024} also leveraged Transformers architecture to developed a sentiment index from South African financial news, specifically using FinBERT with the aim of improving credit risk assessment under International Financial Reporting Standard (IFRS) 9 impairments. The study showed that such a narrative-based index could serve as both a financial signal and a regulatory input, especially in volatile environments.

Later, \textcite{taffler_narrative_2024} examined investor emotions during financial crises and developed crisis-specific emotion dictionaries. They demonstrated that narrative affect extracted from news explained over half of market returns during extreme events, pointing to the central role of emotion in price formation during crises.

The most recent study adopts a more structured approach: \textcite{hong_forecasting_2025} used over 880,000 WSJ articles to forecast U.S. inflation. Using topic decomposition and machine learning models, they showed that narrative predictors significantly outperformed standard macroeconomic models, especially during recessions.

Together, these studies illustrate the evolution of narrative modeling from dictionary-based sentiment measures toward dynamic, data-driven models leveraging deep learning, topic modeling, and Transformers models. Beyond describing narratives, these approaches operationalize them as predictive tools for economic and financial forecasting.


\section{Discussion}
\label{sec:discussions}

This section discusses the key insights derived from the systematic review, structured around the guiding research questions and the patterns identified during the application of the proposed methodology. The discussion is organized into three main dimensions: conceptual, focusing on the definitions and theoretical foundations of narratives; methodological, examining the techniques employed for modeling and analyzing narratives; and practical, addressing the empirical challenges and applications observed in the reviewed studies. Each dimension highlights both the advancements and the limitations within the existing literature, providing a foundation for identifying future research directions.

\subsection{Toward a Shared Understanding of Financial Narratives}

Recent works have increasingly challenged the use of simplistic proxies—such as word frequency counts, aggregate sentiment scores, or isolated topic modeling—for capturing the complexity of narratives. A more holistic perspective is emerging, recognizing that narratives operate as evolving structures combining thematic content, emotional tone, semantic coherence, and temporal dynamics. Under this view, narratives are not static information snapshots but dynamic interpretive frameworks that influence expectations, decision-making, and market behavior.

This evolution in conceptual thinking suggests that modeling financial narratives should move beyond isolated metrics toward more integrated approaches that simultaneously consider multiple dimensions of narrative structure and evolution. By capturing changes in thematic content, semantic meaning, emotional tone, or interpretive coherence, more robust and actionable insights may be extracted, particularly for applications in investment context.

The reviewed literature also indicates an increasing awareness that financial narratives can exhibit heterogeneous effects depending on the assets or markets they pertain to. Understanding how narratives evolve across different asset classes, sectors, or currencies—and how their semantic and emotional characteristics shift over time—could offer more precise and reliable signals than sentiment analysis alone. Such developments point toward a promising research direction where narratives are treated as dynamic, multi-faceted phenomena whose holistic analysis can enhance both explanatory power and predictive capability in financial market applications.

Overall, the emerging consensus supports the need for methodologies that integrate thematic, emotional, and semantic dimensions of narratives, rather than relying on any single proxy. While the operationalization of such holistic models remains an open area of research, this shift lays the foundation for future frameworks capable of capturing the full complexity and market relevance of financial narratives.

\subsection{Methodological Advances in Narrative Quantification}

From a methodological perspective, the field has still progressed from basic sentiment or emotion tagging and topic extraction toward more sophisticated architectures that better capture the complexity of narrative content. This includes unsupervised models for discovering latent narrative structures, deep learning for dynamic sentiment classification, and graph-based methods for understanding narrative coherence and fragmentation over time.

Crucially, these methodological advances coincide with the recognition that narratives operate at different levels—micro (e.g., firm-level disclosures), meso (sectoral or thematic coverage), and macro (aggregate discourse on inflation, risk, or systemic events). Each level requires distinct modeling assumptions and data representations. For example, firm-level narratives often benefit from semantic similarity measures, while macroeconomic narratives lend themselves to topic evolution tracking or network modeling.

The growing use of transformer-based models and domain-specific fine-tuning represents a turning point in narrative modeling. These architectures enable contextual understanding and greater sensitivity to modality, negation, and ambiguity—features that are essential in financial language. However, this increased complexity also raises concerns about interpretability and robustness, particularly when models are used for decision-making or supervision.

The methodological landscape is rich and expanding, but still fragmented. Few studies attempt to benchmark across models or narrative levels. Future work would benefit from methodological synthesis and from clarifying which techniques are most appropriate for different types of narrative inquiry.

\subsection{Narrative Analysis, Behavioral Signals and Market Applications}

Perhaps the most promising insight from the reviewed literature is the empirical validation of narratives as predictive tools. Several studies demonstrate that narrative signals, when carefully extracted, can outperform traditional indicators in forecasting asset prices, returns, volatility, or macroeconomic trends. These results provide strong evidence that narratives are not just epiphenomena but contain economically relevant information.

A notable theme is the connection between narrative features and behavioral finance. Emotional intensity, narrative coherence, and topical salience appear to correlate with phases of market exuberance, uncertainty, and dislocation. In particular, emotionally charged narratives—those expressing excitement, fear, or anxiety—are repeatedly linked to bubbles, crashes, and risk-on/risk-off cycles. This supports the idea that narratives act as vehicles for behavioral contagion, and that narrative monitoring could serve as an early warning signal.

Beyond prediction, narrative modeling is increasingly seen as relevant for systemic risk assessment, regulatory compliance (e.g., IFRS 9 impairments), and investor sentiment analysis. These applications show that narrative analytics is moving beyond academic inquiry toward operational relevance. However, this transition also demands greater transparency and standardization in model design, evaluation, and communication.

One limitation that persists is the reliance on textual sources with unclear representativeness. While surveys, news, and firm disclosures are widely used, they capture different populations and communicative intentions. Understanding the scope and bias of each source remains necessary when translating narrative signals into financial decisions or policy interventions.

In sum, the reviewed literature supports the idea that narratives encode both belief and behavior. They mediate the translation of information into market action and represent a frontier where behavioral finance, macroeconomics, and data science converge.

\subsection{Contribution and Scope}


\section{Conclusion}
\label{sec:conclusion}

This paper introduces a systematic literature review framework that combines algorithmic paper selection with structured data extraction to investigate how financial narratives are conceptualized, quantified, and utilized in financial market analysis. Leveraging NLP and machine learning tools, the review process refined an initial dataset of \nbpapersinitial{} papers down to \nbpapersworkshops{} core contributions, classified into two main categories: narrative understanding and narrative modeling. The effectiveness of the selection framework is supported by the relevance and depth of the insights obtained from the final corpus. The selected research provides key contributions toward addressing the guiding research questions, highlighting the evolving conceptualization of narratives, the methodological advances enabling their modeling, and the potential applications of narrative-based signals in understanding and forecasting financial market behavior.

Studies under narrative understanding primarily seek to define the theoretical, semantic, and rhetorical boundaries of financial narratives. They offer annotation schemes, linguistic typologies, and structural representations that frame narratives as evaluative, temporal, and action-oriented discourses. These works contribute essential groundwork for developing more robust computational models. Secondly, the narrative modeling stream treats narratives as predictive signals. These studies use narrative-derived features to forecast inflation, returns, systemic risk, and investor behavior. They demonstrate that narratives can improve both the explanatory and predictive power of econometric and machine learning models, particularly during periods of market dislocation or heightened uncertainty. Importantly, a few of these contributions go beyond traditional sentiment or topic analysis and incorporate structural and behavioral dimensions, such as narrative virality, emotional polarity, and semantic shift.

The review highlights both the richness and the fragmentation of the current literature. On the one hand, there is increasing recognition that financial narratives are more than expressions of sentiment or thematic frequency: they are structured, evolving, and socially embedded interpretations of economic phenomena. On the other hand, approaches to operationalizing and modeling these narratives remain diverse, with varying assumptions about what constitutes a meaningful narrative signal.

From a methodological standpoint, the field has progressed rapidly, incorporating unsupervised topic modeling, deep learning, and large language models that enabled the construction of narrative indices that reflect attention, sentiment, emotion, and network structure. At the same time, foundational work continues to refine the linguistic, rhetorical, and epistemic dimensions of financial narratives, suggesting the need for continuous improvements in technical approaches. Consequently, this review finds that NLP and textual analysis techniques offer a broad set of tools for quantifying financial narratives in a systematic manner. The evidence strongly suggests that incorporating narrative modeling enhances our understanding of market dynamics, as narrative-based indicators consistently capture behavioral signals that conventional financial and macroeconomic variables often overlook. These indicators not only contribute to improved forecasting performance but also enhance the interpretability of those predictions, providing a structured lens through which to analyze market sentiment, contagion effects, and regime shifts—phenomena that are otherwise difficult to quantify.

In conclusion, the literature reviewed here supports the growing view that narratives are not peripheral to financial markets but are central to how agents form beliefs, process information, and make decisions. As tools for extracting and modeling narratives become more sophisticated, their role in empirical finance, policy analysis, and risk management is likely to expand, offering new insights into the behavioral foundations of economic activity. Nonetheless, current approaches face several limitations: the representativeness of textual data remains uncertain; interpretability of complex NLP models is still a concern; and the validation of narrative indicators across market regimes is limited. Addressing these challenges will be critical to establishing narratives as robust, transparent, and actionable components of financial analysis.


\newpage

\section*{List of Abbreviations}

\begin{itemize}
  \item \textbf{AC} – Agglomerative Clustering
  \item \textbf{API} – Application Programming Interface
  \item \textbf{CN} – Condition Number
  \item \textbf{GMM} – Gaussian Mixture Model
  \item \textbf{IFRS} – International Financial Reporting Standards
  \item \textbf{KMO} – Kaiser–Meyer–Olkin Test
  \item \textbf{LDA} – Latent Dirichlet Allocation
  \item \textbf{ML} – Machine Learning
  \item \textbf{NEG} – Negative Emotion Score
  \item \textbf{NLP} – Natural Language Processing
  \item \textbf{PCA} – Principal Component Analysis
  \item \textbf{SLR(s)} – Systematic Literature Review(s)
\end{itemize}


\newpage

\printbibliography


\newpage

\section*{Appendix}
\label{appendix}

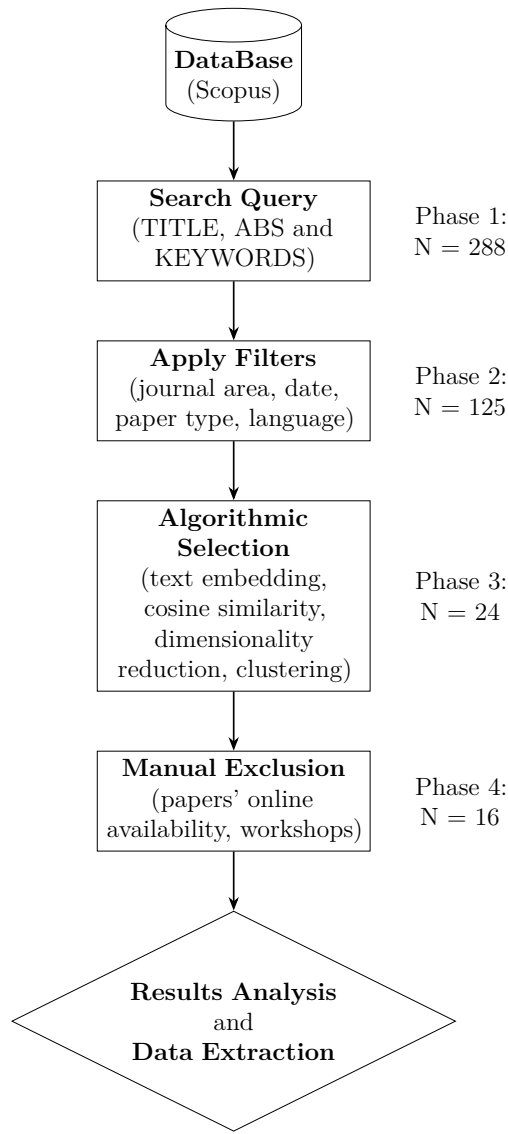
\begin{figure}[ht]
    \centering
    \resizebox{0.45\textwidth}{!}{
        \begin{tikzpicture}[node distance=1cm]
    
            \node (database) [database] {\textbf{DataBase}\\(Scopus)};
            \node (query) [process, below=of database] {\textbf{Search Query}\\(TITLE, ABS and KEYWORDS)};
            \node (n1) [phase, right=0.1cm of query, align=center, text width=2.5cm] {Phase 1:\\N = \nbpapersinitial{}};
            
            \node (filter) [process, below=of query] {\textbf{Apply Filters}\\(journal area, date, paper type, language)};
            \node (n2) [phase, right=0.1cm of filter, align=center, text width=2.5cm] {Phase 2: N = \nbpapersjournalareas{}};
            
            \node (algo) [process, below=of filter] {\textbf{Algorithmic Selection}\\(text embedding, cosine similarity, dimensionality\\reduction, clustering)};
            \node (n3) [phase, right=0.1cm of algo, align=center, text width=2.5cm] {Phase 3: N = \nbpapersalgo{}};
            
            \node (manual) [process, below=of algo] {\textbf{Manual Exclusion}\\(papers’ online availability, workshops)};
            \node (n4) [phase, right=0.1cm of manual, align=center, text width=2.5cm] {Phase 4: N = \nbpapersworkshops{}};
            
            \node (results) [decision, below=of manual, text width=4cm] {\textbf{Results Analysis}\\and\\\textbf{Data Extraction}};
            
            \draw [arrow] (database) -- (query);
            \draw [arrow] (query) -- (filter);
            \draw [arrow] (filter) -- (algo);
            \draw [arrow] (algo) -- (manual);
            \draw [arrow] (manual) -- (results);
        
        \end{tikzpicture}
    }
    \caption{Schematic representation of the paper selection process. The process includes database querying, filtering by inclusion/exclusion criteria, algorithmic selection via NLP and clustering, and manual exclusion of inaccessible studies or workshop proceeds.}
    \label{fig:paper_selection_diagram}
\end{figure}

\begin{landscape}
\begin{longtable}{p{5cm} p{4cm} p{4cm} p{4cm} p{4cm}}
    \caption{Summary of the data extraction phase.}
    \label{tab:data_extraction_table} \\
    \toprule
    \textbf{Paper} & \textbf{Label} & \textbf{Purpose} & \textbf{Method} & \textbf{Narrative} \\
    \midrule
    \endfirsthead
    
    \toprule
    \textbf{Paper} & \textbf{Label} & \textbf{Purpose} & \textbf{Method} & \textbf{Narrative} \\
    \midrule
    \endhead
    
    \bottomrule
    \endfoot

    \textcite{tuckett_tracking_2014} & narrative modeling & Track evolution of emotionally charged financial narratives (phantastic objects) before crises. & Emotion keyword dictionaries, sentiment shift scoring, network analysis of sender-receiver patterns. & Excitement vs. anxiety sentiment index + social network clustering reveal narrative divergence before collapse. \\
    \textcite{hu_annotation_2021} & narrative understanding & Develop a corpus and model for detecting opinion and emotion in financial narratives. & Manual + SpaCy annotation using appraisal theory; dependency parsing, syntactic tagging. & Appraisal-based annotations, intra-sentence pairings of opinion and targets. \\
    \textcite{hsu_narrative_2021} & narrative modeling & Assess how narrative topics in newspaper articles relate to macroeconomic indicators in 1930s China. & Keyword frequency tracking, Ridge/LASSO/Elastic Net regressions, VAR, IRF, Granger causality. & Manual keyword selection and normalized frequency analysis linked to time series regression models. \\
    \textcite{zmandar_cofif_2022} & narrative understanding & Create a large-scale French corpus for summarizing financial reports. & Heuristic + CamemBERT + manual summary extraction. & Summary-level mappings from report sections; NER for entity highlights. \\
    \textcite{chen_covid_2022} & narrative modeling & Assess narrative influence during COVID on market variables using causal testing. & LDA, LM sentiment, Word2Vec, SIR virality, VAR and Granger causality. & Narratives via LDA, semantic shift, and virality scoring, linked to econometric causality. \\
    \textcite{zhu_sentiment_2023} & narrative modeling & Build a future-oriented sentiment index from Weibo posts on housing. & LSTM sentence classification into temporal/sentiment classes, using Word2Vec. & Narrative sentiment split by temporal framing and learned via deep LSTM classifier. \\
    \textcite{sy_fine-grained_2023} & narrative understanding & Improve financial sentiment analysis via argumentative unit detection. & BERT ensemble for argument classification and relation detection. & Pairwise relation classification and claim-premise detection with BERT ensemble. \\
    \textcite{miori_narratives_2023} & narrative modeling & Use GPT + graph theory to extract narrative structure from news and link to market dislocations. & Entity extraction via GPT; co-occurrence graphs; community detection; regress network features on volatility shocks. & GPT-ranked entities + graph metrics (modularity, entropy); topic communities track narrative complexity. \\
    \textcite{borup_quantifying_2023} & narrative modeling & Analyze investor narratives via open-ended surveys during COVID and assess predictive value for markets. & LDA on survey texts; Elastic Net VARs; comparison with media narratives. & LDA-derived narrative topics tracked daily and linked to market behavior via econometric models. \\
    \textcite{stander_news_2024} & narrative modeling & Construct a news sentiment index to act as an early warning for systemic risk and credit impairments. & FinBERT sentiment scoring, PCA on macro indicators, rolling regressions, aspect-based sentiment. & FinBERT sentiment, topic-based aspect analysis, PCA and regressions to link to macro risk. \\
    \textcite{liu_beyond_2024} & narrative understanding & Detect subtle semantic shifts in firm narratives using new financial-STS task. & Triplet-based contrastive learning with GPT-labeled sentence triplets. & Triplet embedding training for narrative similarity under financial framing. \\
    \textcite{agarwal_investor_2024} & narrative modeling & Measure investor emotions from media during market bubbles and link them to returns, volatility, and volume. & Emotion keyword dictionaries; regression on market variables during two Chinese stock bubbles. & Emotion frequency indices (8 emotions), strong vs. weak emotion separation, applied in regression models. \\
    \textcite{taffler_narrative_2024} & narrative modeling & Analyze emotional content of narratives during market crises and its explanatory power over returns and uncertainty. & Keyword-based emotion indices; regression with returns, VIX, volume, EPU during three crises. & Context-specific emotion dictionary applied to financial articles; emotion score time series regressed on market indicators. \\
    \textcite{roos_narratives_2024} & narrative understanding & Review and define the concept of collective economic narratives in economics. & Theoretical review of 436 papers; derive five defining narrative features. & Defines narratives as temporal, socially emergent, sense-making, and action-suggesting story structures. \\
    \textcite{ma_stock_2024} & narrative modeling & Use WSJ-derived narrative index (NEG) to forecast stock returns in the energy sector and beyond. & LDA on WSJ energy topics; predictive regressions; Sharpe/CE utility evaluation. & NEG index from topic attention on WSJ news, tracked as monthly time series and forecast factor. \\
    \textcite{hong_forecasting_2025} & narrative modeling & Forecast U.S. inflation using WSJ-derived narrative features. & LDA for topic modeling, ML regressors (LASSO, ENet, RF, PLS). & LDA narrative topics as ML features in out-of-sample inflation prediction. \\
    
\end{longtable}
\end{landscape}


\end{document}